\documentclass[11pt,twoside]{article}
\usepackage{asp2004}
\usepackage{psfig}
\usepackage{epsf}
\usepackage{graphics}
\usepackage{lscape}
\begin{document}
\title{The Zone of Avoidance: Optical Compared to Near-Infrared Searches}
\author{Ren\'ee C. Kraan-Korteweg} 
\affil{Depto.\ de Astronom\' \i a, Universidad de Guanajuato,
Apdo.~Postal 144, Guanajuato, GTO 36000, M\'exico}
\author{Thomas Jarrett}
\affil{Infrared Processing and Analysis Center, MS 100-22, California 
Institute of Technology, Pasadena, CA 91125}

\begin{abstract}
Galaxies uncovered in the Zone of Avoidance (ZOA) with deep optical
searches are compared to the distribution of objects in the 2MASS
Extended Source Catalog (2MASX). While the galaxy density of optical
surveys is strongly correlated to the dust content, and become
ineffective in uncovering the galaxy distribution at $A_B \ga 3\fm0$,
this effect is much less severe in the NIR. Galaxies can be identified
in 2MASS at optical extinction layers of over $10^{\rm m}$. However, star
density has been found to be the dominant limiting factor in the NIR 
in the wider Galactic Bulge  region (see Fig.~\ref{2M_ZOA}) where
optical surveys still do quite well.
      
Systematic positional offsets have also been found between objects in
the 2MASX and the optical ZOA as well as other galaxy catalogs. These
seem to have their origin in the astrometric reference frame used by
these surveys as well as different fitting algorithms when determining
positions (details are given in the Appendix). The astrometric offsets
between 2MASX and more recent galaxy catalogs (or on the  Digitized
Sky Survey remeasured positions) are of the order of $1\farcs0 -
1\farcs5$, comparable to the relative dispersion in positions between
these surveys.  Still, it is advisable to take this effect into
account when combining galaxies from different catalogs for
observational purposes.
\end{abstract}

\section{ZOA Coverage}
With the final release of the 2MASS Extended Source Catalog (2MASX) by
the Two Micron All Sky Survey Team in 2003 (Jarrett et al. 2000b)
with  over 1.6 million spatially resolved objects, it is now possible
to study the performance of 2MASS in mapping the extragalactic
large-scale structures across the ZOA, as well as compare and
cross-correlate the 2MASS galaxy distribution  with the deep optical
ZOA catalogs. 

In 2000, Kraan-Korteweg \& Lahav summarize that deep optical surveys
have reduced the optical ZOA from extinction levels of $A_B \sim 1^{\rm
m}$  to $A_B \sim 3^{\rm m}$ -- a reduction of about 2.5. Except
for that remaining delimiting gap, which runs parallel along the Galactic
Plane at mostly constant width, this results in a whole-sky galaxy map 
complete to extinction-corrected diameters of  $D^0 \ge 1\farcm3$
(see their Fig. 5).
 
\begin{figure}
\hfil \psfig {figure=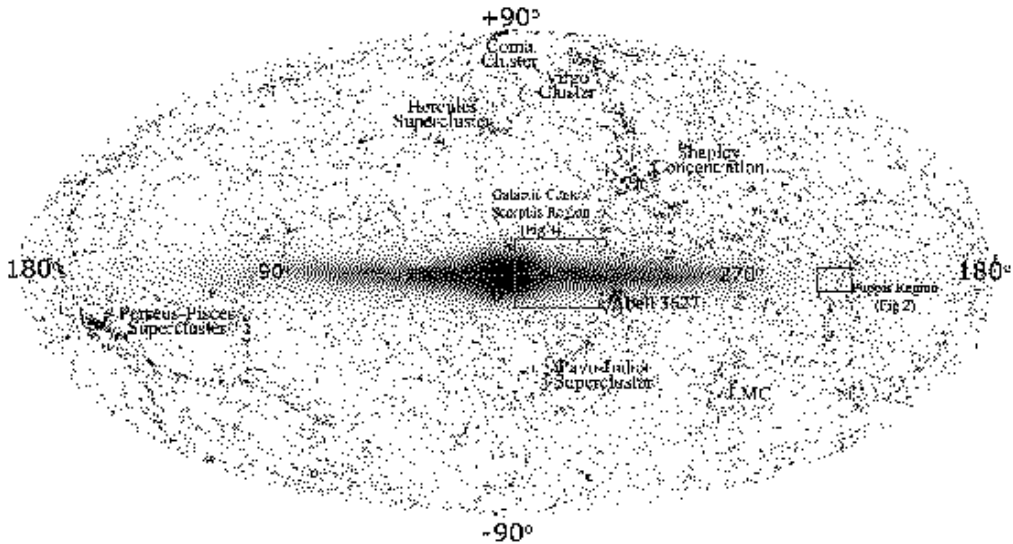,width=13.4cm} \hfil
\caption{Distribution of 2MASX sources in an equal area Aitoff projection
in Galactic coordinates. Stars from the PSC area also displayed.
Prominent large-scale structures are labeled.  The areas that 
are analyzed in this paper are marked with rectangular boxes, i.e.
the Puppis region (Fig.~\ref{6df}) and the Scorpius region (Fig.~\ref{sco}).
The underlying figure comes from Jarrett 2004.}
\label{2MASS}
\end{figure}

2MASX provides a much deeper and uniform view of the whole extragalactic 
with its 1.6 million resolved galaxies (see Fig.~\ref{2MASS})
and an estimated 3 to 5 million unresolved galaxies in the 
2MASS Point Source Catalog (PSC). Whereas 
the galaxy distribution can be traced without hardly any hindrance in
the Galactic Anticenter, the Galactic Bulge -- represented here by 
the half billion Galactic stars from the PSC --
continues to hide a part of the extragalactic sky.   

The  optical and the NIR Zone of Avoidances are not the same. A closer
inspection of both will help to characterize their respective
limitations of uncovering the galaxy distribution behind the  Milky
Way.  By studying the respective magnitude completeness  limits,
colors and surface brightness as a function of extinction  (or star
density -- see  Sect.~2), we hope, amongst others, to optimize
redshift follow-up observations and probe through the Galaxy as deep as
possible.

\begin{figure}
\hfil \psfig {figure=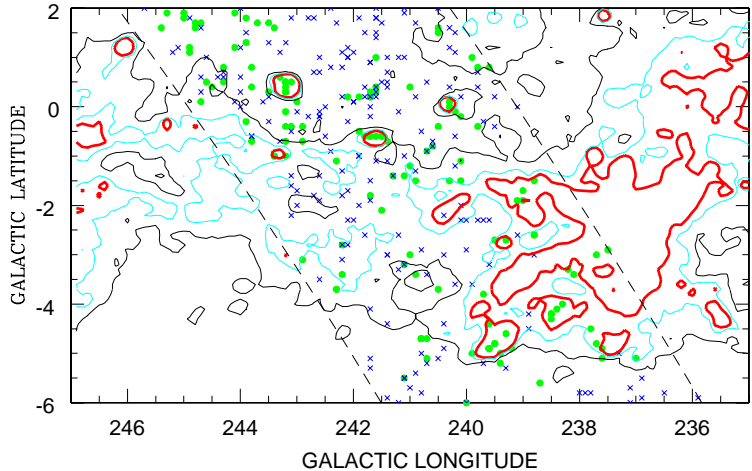,width=10cm} \hfil
\caption{Distribution of 2MASS galaxies observed 
with 6dF in a strip crossing the Galactic Plane in the Puppis region
(see Fig. 1 for the location w/respect to the Milky Way).
Circles represent galaxies with redshifts, crosses without a 
reliable redshift determination. The contours indicate
optical extinction of $A_B=1\fm0, 2\fm0$ and $3\fm0$ 
(thick contour) as given in the DIRBE/IRAS reddening maps
(Schlegel, Finkbeiner, \& Davis 1998).}
\label{6df}
\end{figure}

A reduction of the ZOA on the sky does not imply at all 
that a similar reduction can be attained in redshift space. This is 
seen quite clearly in Fig.~\ref{6df} which shows 2MASS 
galaxies in Puppis -- a filament is crossing the Plane there (see 
Fig.~\ref{2MASS}) -- on a sequence of 6-degree fields (6dF) centered on Dec 
$= -25\deg$. This strip has been observed with the multifibre spectroscope 
at the UK Schmidt Telescope as part of a pilot project aimed at 
extending the 6dF Galaxy Survey (see http://www.mso.anu.edu.au/6dFGS 
for further details) towards lower latitudes (it now is restricted 
to the southern sky with $|b| \ge 10\deg$).

The crosses mark galaxies for which a reliable redshift could be
measured, the filled dots galaxies for which this could not be
realized. At first glance, the plot seems to indicate a fantastic
success rate for obtaining redshifts all the way across the Milky Way.
However, it should be noted that the concerned ZOA strip lies in a
region reknown for its low dust content (it hardly exceeds 3
magnitudes), and a careful inspection indicates that redshifts have
generally not been obtained for  $A_B \ga 3\fm0$ (thick
contour).  So even when galaxies are identifiable deep in the Plane,
reducing the {\sl redshift ZOA} will remain hard,  and an
optimization of targeted ZOA galaxies is crucial for redshift
follow-ups.



\begin{figure}[!ht]
\hfil \psfig {figure=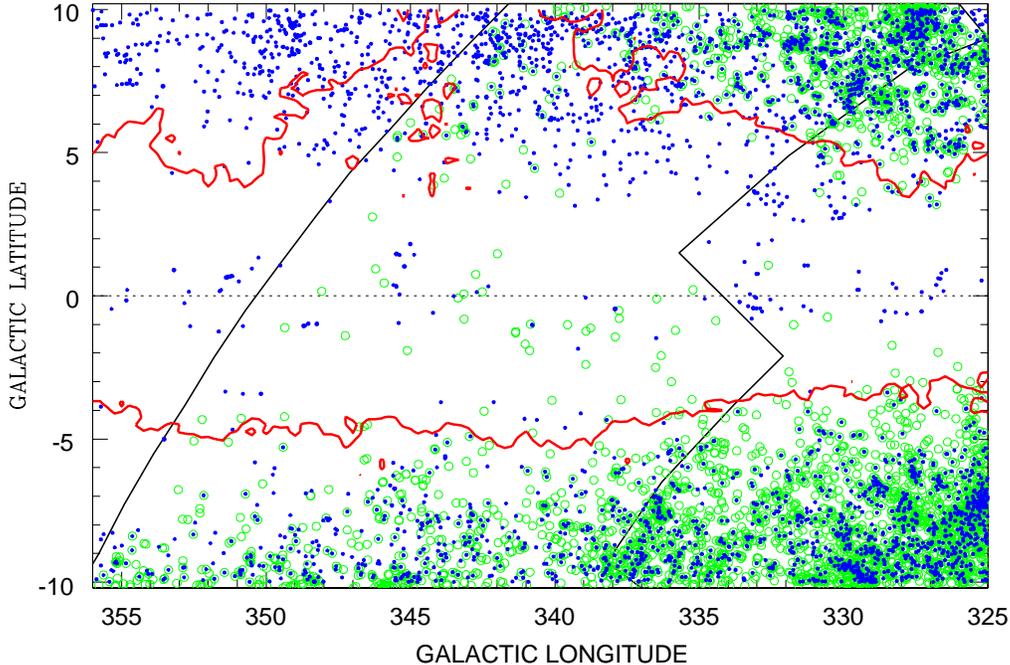,width=13.4cm} \hfil
\caption{Distribution of optically detected galaxies (circles) with 
$D > 12\arcsec$ in the Scorpius and adjacent Galactic Center region
(see Fig. 1 for the location w/respect to the Milky Way),
and 2MASS galaxies (small dots) with $K \le 13\fm5$ 
and $(J-H) > 0\fm0$. The contour marks an extinction level of $A_B=3\fm0$ 
(Schlegel et al. 1998).}
\label{sco}
\end{figure}

With the preparation for publication of the fourth of our deep optical
galaxy surveys in Scorpius (Fairall \& Kraan-Korteweg, in prep.;  see
Fairall \& Kraan-Korteweg 2000 for preliminary results), we compared
the optical galaxy distribution with 2MASX objects in the Scorpius
region (in Fig.~\ref{2MASS} this regions lies just to the right of the
Galactic Center, where confusion from the foreground Milky Way is
extreme). Both the optical and 2MASS galaxy distributions are
displayed in more detail in Fig.~\ref{sco}.  There the large circles
represent optically detected galaxies in Scorpius and in the adjacent
Great Attractor (GA) region (Woudt \& Kraan-Korteweg 2001). The small
dots represent the 2MASX objects  with $K \le 13\fm5$, the
completeness limit at high Galactic latitudes, excluding extremely
blue objects with $(J-H) < 0\fm0$ which are merged star  images
(Jarrett et al. 2000a). It should be noted that this plot represents
{\sl all} the objects of 2MASX, hence also contains a relatively small
number of Galactic extended objects  such as HII regions and planetary
nebulae, the prime contaminant at $|b| < 2\deg$ (optical catalogs 
also contain a small fraction of them). The 3\fm0 optical
extinction contour is also drawn, i.e. the level to which deep optical
surveys  are fair tracers of the galaxy distribution (Kraan-Korteweg
2000; Kraan-Korteweg \& Lahav 2000).

At positive latitudes, an analysis of the distribution confirms that
deep optical searches are faithful tracers of the galaxy distribution
to $A_B \sim 3^{\rm m}$ with only a few galaxy candidates peaking
through higher dust levels, whereas the 2MASX sources seem to probe
deeper into the Milky Way. However, at negative Galactic latitudes this trend
is not as clear cut. There it appears that more optical galaxies
are found than 2MASS galaxies and that the optical searches find
galaxies closer to the Plane than 
the 2MASS catalog of resolved objects (Jarrett et al. 2000a), an effect 
also seen in the adjacent GA region. 
A direct correlation between dust absorption and 2MASX 
source density is not seen here.

\begin{figure}
\hfil \psfig {figure=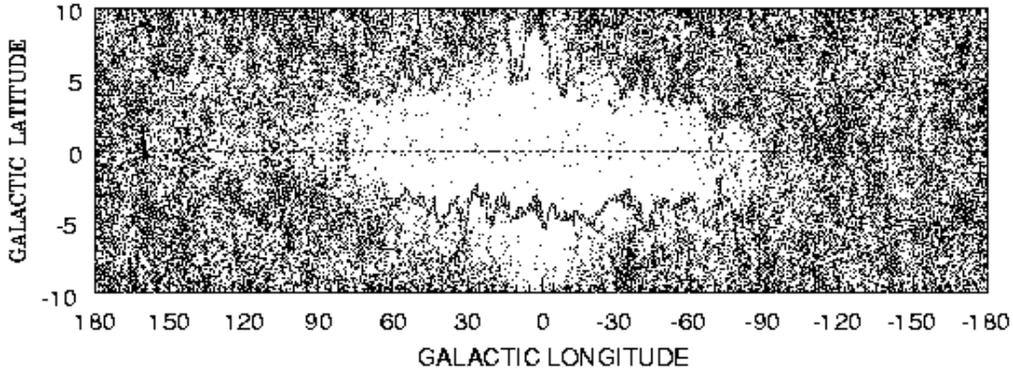,width=13.4cm} \hfil
\caption{Distribution of 2MASS galaxies with $K \le 13\fm5$ 
and $(J-H) > 0\fm0$. The contour marks an optical extinction level 
of $A_B=3\fm0$.}
\label{2M_ZOA}
\end{figure}

This can be verified in Fig.~\ref{2M_ZOA} where the extinction
contour of $A_B = 3\fm0$ is superimposed on the distribution of 2MASX
sources  for the whole Milky Way ($|b| \le 10\deg$). It is undeniable 
that extinction and source density are not directly related. In the Galactic
Anticenter (AC), roughly defined as $\ell: 180\deg \pm 90\deg$ here, 2MASX
objects seem to cross the Plane without any hindrance (see 
the allsky plots from Jarrett 2004). In fact, in
plots of NIR  magnitude or diameter versus $E(B-V)$ (not shown here), 
it is clear that galaxies can be easily identified up to extinction 
levels equivalent to $A_B \ga 10^{\rm m}$, whereas in plots of 
$K^o$-band magnitudes versus $E(B-V)$ imply that 
2MASS remains quite complete up to $K^o \la 13\fm0$ for $E(B-V) \sim 
2\fm6$. This is not at all true for the wider Galactic Bulge (GB) region,
where 2MASS uncovers objects to much lower 
extinction levels and is at least one magnitude less deep compared to 
the AC.  

The obvious culprit here is the star density which increases strongly
towards the Galactic Center. This progressive loss of galaxies towards
the GB is already noticeable when simply comparing the fraction of
galaxies that have 2MASS counterparts in the optical catalogs, which
decreases from 47\% in the Hydra/Antlia (H/A) region (Kraan-Korteweg 2000), 
to 39\% in  Crux, 37\% in the GA (Woudt \& Kraan-Korteweg) and
33\% in the Scorpius region (Fairall \& Kraan-Korteweg, in prep.). In
fact a near perfect imprint of the Galactic Bulge is given in 
Fig.~\ref{2M_ZOA} by the lack of objects in this area.

A comparison of the optical and 2MASS galaxies over the whole region of the
four deep optical catalogs ($270\deg < \ell < 350\deg, |b| < 10\deg$)
finds the optical galaxian distribution denser and more clumped. In a
comparison of magnitude, diameter and colors versus extinction of  the
galaxies common to both the optical and NIR catalogs, it is obvious
that the optical surveys miss galaxies with $D>22\arcsec$ for $A_B \ga
2\fm5$ and at 4 magnitudes of extinction become incomplete for $D \ga
50\arcsec$ galaxies, which is quite consistent with the findings in
Kraan-Korteweg 2000, and Woudt \& Kraan-Korteweg 2001. As expected
 the majority of the 2MASS galaxies not recovered in the
optical searches are redder and faint (mostly with $K \ga 12^{\rm
m}$). This is partly a ZOA effect with NIR surveys finding
red galaxies more readily at higher absorption level. But it is 
also due to the inherent characteristics of the two surveys. Whereas 
the NIR is better at detecting old galaxies and ellipticals, the optical 
surveys -- although susceptible to all galaxy types -- are best at finding
spirals and late-type galaxies (especially LSB galaxies and dwarfs).
The surveys are in fact complementary.
 
These results suggest that an optimal approach to unveiling the
mask of the Milky Way, revealing the background galaxy distribution, is to
combine deep optical surveys with those of the near-infrared.  The former
are sensitive to galaxies located behind regions of high source confusion,
and the latter aresensitive to galaxies located behind thick dust walls.

\section{Systematical Positional Differences}
While preparing a field in the ZOA in Scorpius for observations  with
the 6dF multifibre spectroscope at the UK Schmidt Telescope, optical
and 2MASS galaxy samples were combined. It was then noted by  W. Saunders
(priv.~comm.) that the optical positions had a systematic shift in
RA  of approximately $1\arcsec - 2\arcsec$ compared to the 2MASS
positions. The  latter are based on the intensity-weighted centroid of
the combined  $J H K$ co-added image and are of high accuracy
($0\farcs3$; Jarrett et al. 2000b), whereas  the optical positions for
the Scorpius galaxies were measured on  the 2nd Generation Digitized
Sky Survey images (DSS2) by centering the cursor by eye on the galaxy
(Fairall \& Kraan-Korteweg,  in prep.). So we did expect a scatter of
the order of about $1\arcsec$, but  a systematic offset of that
order seemed inexplicable.

\begin{figure}
\hfil \psfig {figure=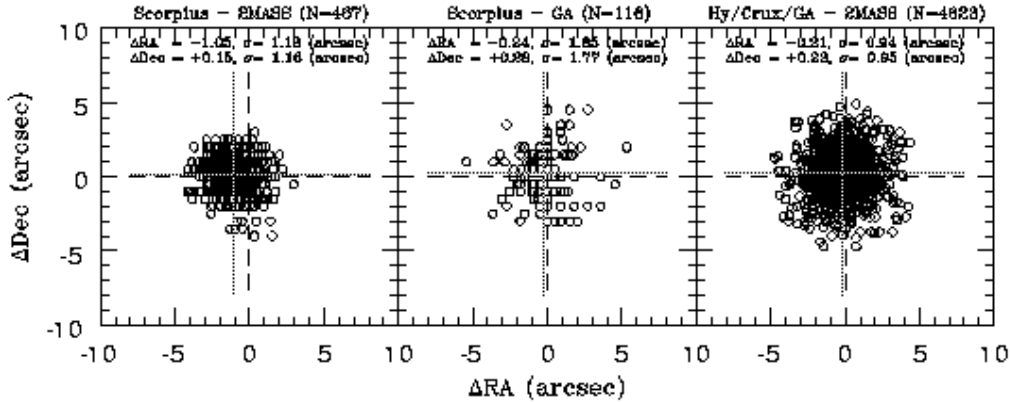,width=13.4cm} \hfil
\caption{Comparison of positions in the ZOA: optical galaxies  in common 
with 2MASS in Scorpius (left panel); optical galaxies detected in both 
the Scorpius and GA optical surveys (middle panel); optical galaxies 
in common with 2MASS in the H/A, Crux and GA searches.}
\label{dr_dd}
\end{figure}

We first investigated this in further detail for the Scorpius  region
itself (see left panel of Fig.~\ref{dr_dd}). For the 463 (33\%)
galaxies in common in the Scorpius region,  the mean shift in Right
Ascension could indeed be verified  and quantified at  $\Delta{\rm RA}
= -1\farcs05, \sigma = 1\farcs18$,  with a smaller offset in
Declination, $\Delta{\rm Dec} = +0\farcs15, \sigma = 1\farcs16$.

Not understanding this offset, we then compared the positions of
Scorpius galaxies in common with the GA in their overlap region
(N=116). The latter were measured directly from the ESO/SERC IIIaJ
film copies with the Optronics high precision measuring machine of ESO
(Woudt \& Kraan-Korteweg 2001) and  proven to have an accuracy of $
\la 1\arcsec$  (Kraan-Korteweg 2000).  The Scorpius shift could not be
recovered. The dispersion is quite large here (middle panel of
Fig.~\ref{dr_dd}). Considering that overlap galaxies are  always
located  on the edges of the sky survey plates, and thus have the
largest errors, this is not surprising.

As a further resort in unraveling the origins of these systematics, we
then plotted the differences in position of all the galaxies in the
combined ZOA catalogs H/A, Crux, and GA  -- all measured with
Optronics  -- versus their 2MASS counterparts (right panel of
Fig.~\ref{dr_dd}).  Contrary to expectations, the mean offset was
found to be small ($\Delta{\rm RA} = -0\farcs22, \sigma = 0\farcs94$,
$\Delta{\rm Dec} = +0\farcs24, \sigma = 0\farcs95$), and of the same
order as the Scorpius/GA comparison.

\begin{figure}[t]
\hfil \psfig {figure=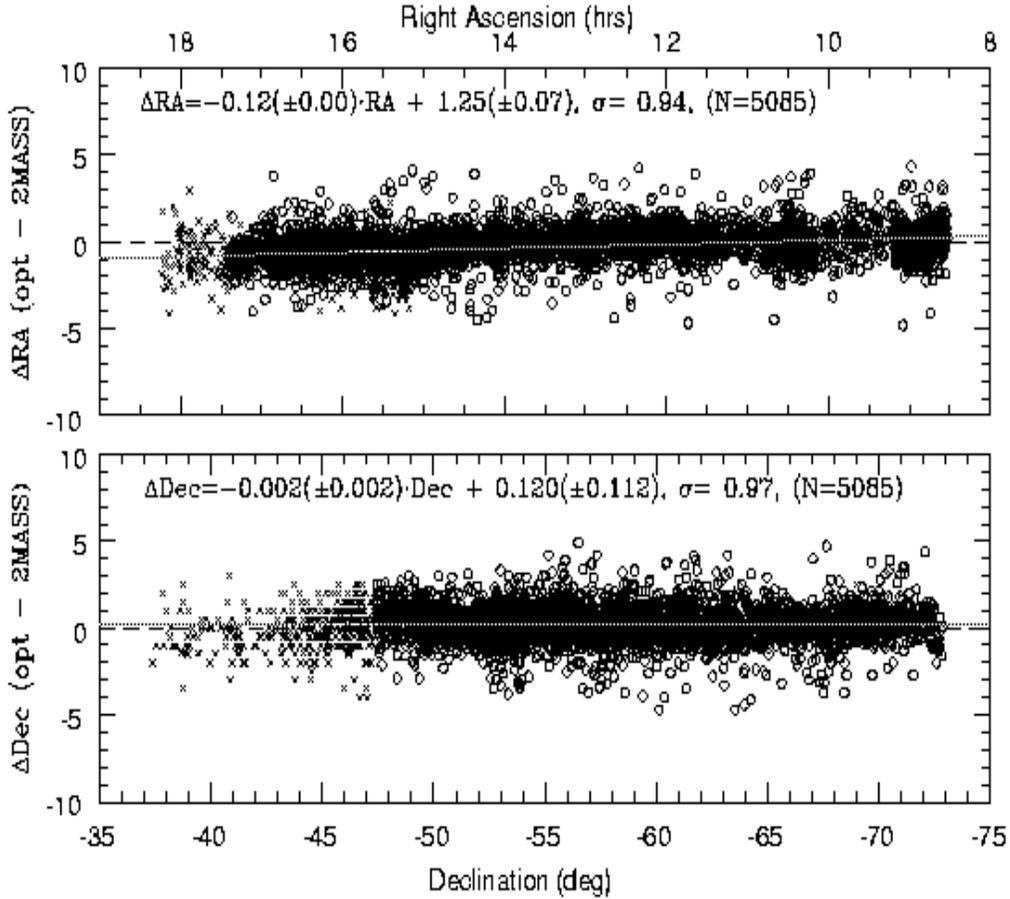,width=13.4cm,height=12cm} \hfil
\caption{Positional offsets as a function RA (top panel) 
and Dec (bottom panel) between the optical ZOA surveys 
and 2MASS. Circles: galaxies measured with Optronics in 
the Hy/Ant, Crux and GA regions; crosses: galaxies  
measured on DSS2 in Scorpius.}
\label{ra_dr}
\end{figure}

A logical connection between these three discrepant results could only
be made when the positional differences  $\Delta{\rm RA}$ and
$\Delta{\rm Dec}$  were plotted as a function of RA and Dec. As
Fig.~\ref{ra_dr} illustrates, a {\sl dependence} of the  RA-offset as
a function of RA exists ($\Delta {\rm RA} = -0.12 {\rm RA(^{\rm h})} +
1\farcs25$). This trend, defined by the H/A, Crux and GA galaxy
positions is followed to the dot  by the  Scorpius galaxies.  The
linear relation explains the offset of about $-1\arcsec$ since 
it corresponds to the mean of the RA-range of the Scorpius galaxies 
($15^{\rm h}$ to $18^{\rm h}$).

In Declination there is no significant trend. The mean
offset of all 5085 common ZOA galaxies is 
$\Delta {\rm Dec}=+0\farcs23,
\sigma=0\farcs97$.  

This does not, however, explain the origin of the offsets between
2MASX, the optical ZOA catalogs and DSS. In order to find whether
other galaxy catalogs show similar offsets and trends, we investigated
this in further detail using the 2MASS extended sources  of the
Winter 1999 public data release by Jarrett.  The details of  this
analysis are given in the Appendix. It suffices to summarize here
that clear deviations have also been found between 2MASS and
other optical  galaxy positions. They vary from one catalog to
another, and all show displacements between the northern and southern
hemispheres  (see Fig.~\ref{dr_dd_tj}). At the root of these
systematics lie the different astrometric reference frames with which
the galaxy  catalogs are calibrated (see Table~\ref{astrom}). Many rely
on, or are tied  to, old star catalogs which in addition are valid
for different sections of the sky. The  second source of the
discrepancies have their origin in the digitization of sky
surveys. Many modern galaxy catalogs list positions that are linked to
star catalogs that were made from digitized sky surveys (such as the
Guide Star Catalog, GSC; Lasker et al. 1990) and/or are  determined on 
digitized sky images. Here the fitting algorithms as well as the pixel 
size of the digitization determine the zeropoints of the catalogs and 
limit the accuracy  with which positions can be derived.

\section{Discussion}

\subsection{Penetration of the ZOA}
With regard to penetrating the ZOA, optical galaxy catalogs are mainly
limited by Galactic extinction. The deep ZOA searches have reduced the
optical ZOA considerably, but become increasingly ineffective at
extinction levels of over $A_B \sim 3\fm0$ (see  Kraan-Korteweg \&
Lahav 2000).

In the NIR, with ($A_K = 0.09 \cdot A_B$), dust obscuration is a much
weaker  effect. In plots of diameter or NIR magnitudes versus
$E(B-V)$, the effect of absorption becomes less stringent as one 
moves to redder wavebands. However, a strong delimiting factor in 
ZOA-penetration in the NIR is the star density. As shown with  
Fig.~\ref{2M_ZOA}, 2MASS does not uncover galaxies at low Galactic 
latitudes in the Galactic Bulge region ($\ell \la \pm 90\deg$), an   
effect completely uncorrelated with extinction, but is due to star crowding
and the underlying stellar confusion noise (an effect that is 
amplified by the relatively large ($2\farcs5 - 3\arcsec$) point 
spread function of the 2MASS survey).


This fact is confirmed when plotting {\it extinction-corrected}
$K^o$-band magnitudes versus $E(B-V)$ for the Galactic Bulge (GB)  and
Anticenter (AC) region. Whereas 2MASS remains quite complete  for $K^o
= 13\fm0$ up to extinction levels of $E(B-V) \sim 2\fm6$  in the AC,
it does not uncover galaxies to that extinction level  in the GB and
seems overall less deep (at least one magnitude)  compared to the AC.

Deep optical ZOA galaxy searches thus are actually very useful  in the
GB region as they can uncover galaxies to lower extinction
levels and higher star density levels than 2MASS. It should be
maintained, however, that for the remaining ZOA -- respectively the
whole sky -- 2MASS as a homogeneous whole-sky survey obviously is  far
superior, particularly considering that the optical ZOA catalogs
were not only performed on different plate material, but also by
different searchers using different search techniques.

\subsection{Positional Accuracy}
A careful comparison of 2MASS positions with various other catalogs
reveal systematic differences, which differ between the northern and
southern hemisphere as well as from one catalog to another. The
offsets and trends seem  to have its origin in the different
astrometric  reference frames as well as the software routines with
which positions  were determined.

A dependence in offset as a function of position, as seen for the ZOA
optical catalogs (H/A, Crux, GA and Scorpius, see Fig.~\ref{ra_dr})
has not been seen in other catalogs. Then again, the galaxies and 
sky coverage here explored did extend over a limited RA and Dec
ranges ($10^{\rm h} -  14^{\rm h}$). They did not include the RA range 
where this effect becomes significant ($15^{\rm h} -  18^{\rm h}$). 
Hence it remains unclear whether this dependence is inherent to other 
optical catalogs as well. This certainly merits further investigation.
 
One might argue that future catalogs should all standardize their
positions to the Tycho/Hipparcos reference frame (or the highly
accurate 2MASS and FIRST positions) to avoid such systematical
offsets. However, as long as DSS images (first or second generation)
are used for position measurements or multiwavelength sky overlays,
these offsets will not disappear, because this survey material is tied
to the old astrometric star catalogs. Moreover, the $rms$ will remain
of the order of $1\arcsec  - 1\farcs5$ because of the pixel size of
the DSS.

The systematic differences that exist between more modern catalogs are
on average small (of the order of $1\farcs0 - 1\farcs5$) and should not
cause problems except when doing precision spectroscopy. However,
care should be taken when different catalogs are combined for instance for
multi-object  fibre spectroscopy. If the fibers intercept only small
areas on the sky, the systematic offsets between catalogs may lead to
un-optimized positioning  of the fibres on the galaxy, reducing the
expected or required signal-to-noise ratio or missing a small galaxy
altogether.

\acknowledgements   We kindly thank Baerbel Koribalski for her helpful
suggestions. This research has made use of the NASA/IPAC Infrared
Science Archive (2MASS) and the NASA/IPAC Extragalactic Database
(NED), which are operated by the Jet Propulsion Laboratory, California
Institute of Technology, under contract with the National Aeronautics
and Space Administration. RCKK thanks CONACyT for their support
(research grant 40094F) and the Australian Telescope National
Facility (CSIRO) for their hospitality during her sabbatical.

\section*{Appendix} 
\subsection*{NED versus 2MASS Positions}

The 2MASS extended sources of the  Winter 1999 public data release by
Jarrett (see
http://spider.ipac.caltech.edu/staff/jarrett/2mass/XSC/astrometry)
were used to research the possible systematics of
positional offsets. As before the galaxies were divided into  northern
and southern hemisphere, including only those galaxies for
which  the NASA Extragalactic Database (NED) reported accurate
coordinate positions (uncertainty major  axis $<2\arcsec$). The
galaxies all lie within Right Ascension range of $10^{\rm h} < {\rm
RA} < 14^{\rm h}$.

\begin{figure}
\hfil \psfig {figure=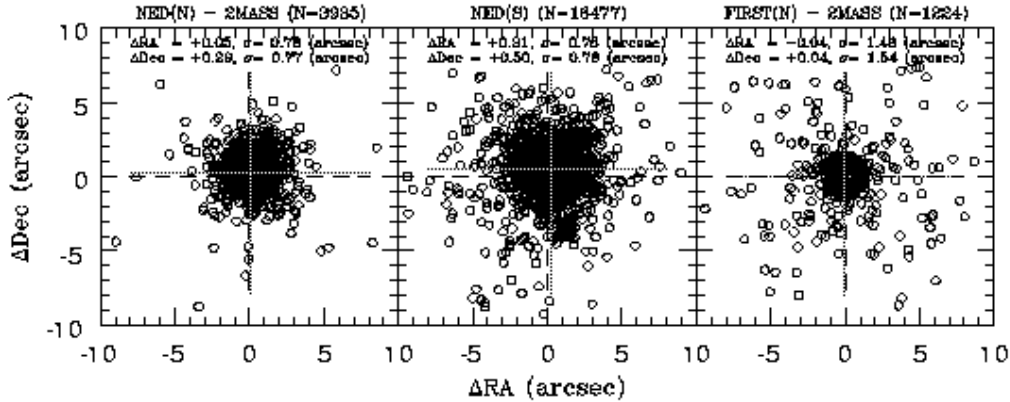,width=13.4cm} \hfil
\caption{Comparison of 2MASS positions with optical galaxies in 
the northern hemisphere (left panel), optical galaxies in the south 
(middle panel) and with radio galaxies (FIRST).}
\label{dr_dd_tj}
\end{figure}

In the north, this led to 3935 counterparts that are classified in NED
as galaxy (G) or infrared source (IrS). As shown in the left
panel of Fig.~\ref{dr_dd_tj} the mean  offsets with  $\Delta{\rm RA} =
+0\farcs05, \sigma = 0\farcs76$ and  $\Delta{\rm Dec} = +0\farcs29,
\sigma = 0\farcs77$ are small.

The result is based on a mixture of over a dozen galaxy catalogs, many
of which not only show discrepancies from one catalog to the
next but also shifts or large dispersions within the catalogs
themselves. The Abell cluster galaxies (Abell 1958, Abell Corwin, 
\& Olowin, 1989), for instance, display mean displacements of up to 
$\pm 5\arcsec$ between different clusters.
Excluding these catalogs does not really alter the mean result for the 
north though. The mean shift of all of them together is the same 
as for two dominant catalogs, i.e. the Lick Northern Proper Motion 
Galaxy  Catalog (NPM1G;
Klemola, Jones, \& Hanson 1987), and the North Galactic Pole Survey
(NGP9; Odewahn \& Aldering, 1995), both of which have consistent
shifts (Table~\ref{astrom}).

In the south, the number of 2MASX coincidences is considerably higher,
mainly due to the overlap with the APM galaxy catalog (APMUKS; Maddox
et al. 1990) and  the Las Campanas Redshift Survey (LCRS; Shectman et
al. 1996) with  8189 and 6051 galaxies each. It is obvious from
Fig.~\ref{dr_dd_tj} that the positional offsets with  $\Delta{\rm RA}
= +0\farcs31, \sigma = 0\farcs76$,  $\Delta{\rm Dec} = +0\farcs50,
\sigma = 0\farcs76$, are distinctly larger compared to the north.
There exists some variation from one galaxy catalog to the next (and
even from one LCRS Declination strip to the next) but the offsets in
the south all are consistently higher than in the north, independent of
how we subdivide the sample (see also Table~\ref{astrom}). This is
substantiated when comparing northern and southern galaxies within
individual catalogs such as the NGC (Dreyer 1888), MGC
(Vorontsov-Velyaminov, Archipova, \& Krasnogorskaja 1962-1975), IRAS
(IRAS PSC; Joint IRAS Science Working Group 1988) or the Zwicky et
al. (1961-1968) CGCG galaxies. In all of them, the discrepancy
between north and south persists. This can also be seen in Table 1
for the more recent  NPM1G galaxies ($\Delta{\rm RA} = 0\farcs05$ 
in the north vs. $0\farcs41$ in the south and $\Delta{\rm Dec} = 
0\farcs28$ vs. 0\farcs68 respectively).
 
The dependency of the offset as a
function of RA  (Fig.~\ref{ra_dr}) is not detected in any of the NED
subsamples (except for the lower accuracy positions LCRS galaxies;
there the trend is even stronger than for the ZOA galaxies).
Then again the shift for the ZOA galaxies becomes significant 
only at the higher RA range (RA$ \ga 14^{\rm h}$). Indeed, 
the ZOA galaxies would not have revealed this dependency if the
study would have been restricted to $10^{\rm h} - 14^{\rm h}$. 

The only good agreement without any systematics 
could be seen in a comparison of 2MASX objects with radiosources from 
the FIRST Survey (Becker, White, \& Helfand 1995). As illustrated 
in the right panel of Fig.~\ref{dr_dd_tj}, the mean offsets are 
insignificant. This result was confirmed in August 2002 with 
the then available 23,000 matches between 2MASS and FIRST within a
5\arcsec\ radius 
(http://spider.ipac.caltech.edu/staff/jarrett/2mass/XSC/jarrett\_FIRST.htm).

\subsection*{Origin of Deviations}

No differences between 2MASS positions and radio positions exist,
while clear deviations occur between 2MASS and optical galaxy
positions. They may vary from one catalog  to another, and all
systematically show shifts between  the northern and southern
hemispheres. What lies at the origin of these discrepancies? The
culprit seems to be the astrometric reference frames.

FIRST claims positional accuracy at the 90\% confidence level of
$<1\arcsec$. Its positions are linked to the VLB reference frame and
should have absolute astrometric offsets of  $<0\farcs15$ (Becker et
al. 1995).  The 2MASX intensity weighted centroid positions from the
combined $J H K$ images are claimed to have an estimated $rms$
uncertainty of $0\farcs3$  (Jarrett et al. 2000b) and are tied to the
Tycho Catalog (ESA 1997). 
According to  H{\o}g et al. (1997) the latter conform to the
ICRS (International Celestial Reference System) to better than 1~mas
with median astrometric errors of about 25~mas. The
correspondence between two catalogs confirm their high  
accurate absolute astrometry. 

\begin{table}[!t]
\caption{Systematic offsets of 2MASS positions with NED and ZOA galaxies}
\smallskip
{\small
\begin{center}
\begin{tabular}{lcrrrrrl}
\tableline
\noalign{\smallskip}
   hemisphere & catalog & N     & $\Delta {\rm RA}$ & $\sigma $  & $\Delta {\rm Dec}$ & $\sigma $ & astrometric \\
              &         &       & (\arcsec)         & (\arcsec)  &
(\arcsec)          & (\arcsec) & ref.~frame \\
\noalign{\smallskip}
\tableline
\noalign{\smallskip}
   north (opt)     & various &  3935 &  0.05  & 0.76 & 0.29    & 0.77 & various \\
              & NGP9    &   344 &  0.00  & 1.05 & 0.29    & 0.87 & AGK3+\\
              & NPM1G   &  1426 &  0.05  & 0.34 & 0.28    & 0.35 & AGK3     \\
\noalign{\smallskip}
\tableline
\noalign{\smallskip}
south (opt.)        & various & 16477 &  0.31  & 0.76 & 0.50    & 0.76 & various \\
              & APMUKS  &  8196 &  0.34  & 0.83 & 0.57    & 0.64 & Perth 70  \\
              & NPM1G   &  1222 &  0.41  & 0.66 & 0.68    & 0.64 &
Perth 70     \\
              & LCRS    &  6051 &  0.37  & 0.61 & 0.48    & 0.60 & GSC     \\
\noalign{\smallskip}
              & H/A-Crux-GA &  4623 & -0.24  & 0.94 & 0.28    & 0.95 & PPM     \\
              & $10{\rm h}-14{\rm h}$ only &  1714 & 0.05  & 0.90 & 0.21    & 0.98 & PPM     \\
              & FKK     &   463 & -1.05  & 1.18 & 0.15    & 1.16 & DSS2  \\
\noalign{\smallskip}
\tableline
\noalign{\smallskip}
north (radio)         & FIRST   &  1224 & -0.04  & 1.45 & 0.04
& 1.54 & VLA \\
\noalign{\smallskip}
\tableline
\end{tabular}
\end{center}
AGK3+: AGK plus secondary stars
}
\label{astrom}
\end{table}

The systematic deviations found between 2MASX and other optical
catalogs must thus have the root in their astrometry.  
In order to verify that, we had a look at the astrometry 
reference frames of the comparisons presented here (see last 
column in Table~\ref{astrom}). These are:
\begin{center}
{\small
\begin{tabular}{lcl}
\tableline
\noalign{\smallskip}
AGK3 & $ 2\deg < \delta < 90\deg$ & Heckman et al. 1975\\
SAOC & $ -65\deg < \delta < 2\deg$ & SAO 1966 \\
CPC  & $ -90 < \delta < -65\deg$ & Stoy 1966, 1968 \\
Perth 70 & south & H{\o}g \& van der Heide 1976\\
PPM  & all-sky & Roeser \& Bastian 1993 \\
GSC  & all-sky & Lasker et al. 1990 \\
DSS 1 \& 2  & all-sky & GSC  \\
\noalign{\smallskip}
\tableline
\end{tabular}
}
\end{center}
Considering the variety in star catalogs that are at the root of
the astrometric calibration, the systematic offsets between
different galaxy catalogs seem understandable. But does this
explain (a) the similarities in the shifts of catalogs with different
astrometry, and (b) the consistently found discrepancy in offsets 
between hemispheres? 

It actually does. The reason is that most of the astrometric calibrations
have their roots in three main astrometric reference catalogs,
i.e. AGK3, SAOC and CPC, each covering a different declination
range. This might explain some of the consistent differences 
between north and south. 

Moreover, many of the later reference 
frames are tied to these early star catalogs, such as the Guide 
Star Catalog as well as DSS1 and DSS2. 
The astrometry of the GSC relies on the AGK3, SAOC and CPC and digitized
images of pixel size was 1\farcs7. Therefore the individual positions 
are not very accurate  ($\sim1\arcsec$) and might suffer systematic 
errors of up to several  arcsec close to the Schmidt plate-edges 
(V\'eron-Cetty  \& V\'eron 1996; see also Deutsch E.W. 1999 for a 
more detailed discussion on the astrometry error using digital sky surveys).
The same inaccuracies are inherent to any position derived form digitized
sky surveys such as the DSS1 or DSS2 (the pixel size of 1\farcs7 
and 1\farcs0 respectively).


So the systematic shifts between different galaxy catalogs, as well
as north and south seem to have their origin in the different 
astrometric reference frames, whereas many of the 
similarities in offsets are found for catalogs either based on 
the same astrometric reference frames or on secondary frames
which again are linked to the original three astrometry frames.
One should thus always be aware when determining positions from 
digitized images -- even if calibrated to highly accurate  
astrographic catalogs -- that the precision one can attain does 
not depend on that catalog, but on the pixel size of the scanning 
machine that produced the digitized images.

On top of that, Corwin et al.~1998 noted that further systematics 
of the same size may be introduced through the application of 
different position fitting algorithms, depending on whether the 
routines center or corners of pixels as their zeropoint 
(the point of origin of a DSS scan is, for instance, the
southeastern corner of the southeastern pixel and not the center 
of the pixel) or format differences (integer versus floating).

\end{document}